\documentclass[letterpaper]{raa}            %

\usepackage{natbib}  
\usepackage{graphicx,times,amssymb}             
\input{epsf.sty}                        
\input{psfig.sty}                       
\usepackage[totalwidth=480pt,totalheight=680pt]{geometry}
\usepackage{multirow}

\usepackage{enumerate}

\def\cm{\,{\rm cm}}

\def\ergscm2 {erg\,s$^{-1}$cm$^{-2}$}

\def\cm2 {cm$^{-2}$}

\def\aap {A\&A}
\def\apj {ApJ}
\def\apjs {ApJS}

\def\mnras {MNRAS}

\def\nat{Nature}

\begin{document}

\title{SN 2009ip and SN 2010mc as dual-shock  Quark-Novae}

 \volnopage{Vol.0 (200x) No.0, 000--000}      
   \setcounter{page}{1}          

   \author{Rachid Ouyed
      \inst{1}
   \and Denis Leahy
      \inst{1}
   }

\author{Rachid Ouyed, Nico Koning and Denis Leahy}

 \institute{Department of Physics and Astronomy, University of Calgary, 
2500 University Drive NW, Calgary, Alberta, T2N 1N4 Canada; {\it rouyed@ucalgary.ca}}

\abstract{In recent years a number of double-humped supernovae have been discovered.  This is a feature predicted by the dual-shock Quark-Nova model where a SN explosion is followed (a few days to a few weeks later) by a Quark-Nova explosion.  SN 2009ip and SN 2010mc are the best observed examples of double-humped SNe. Here, we show that the dual-shock Quark-Nova model naturally explains their lightcurves including the late time emission, which we attribute to the interaction between the mixed SN and QN ejecta and the surrounding CSM. Our model applies to any star (O-stars, LBVs,  WRs etc.) provided that the SN explosion mass is $\sim 20M_{\odot}$ which point to the conditions for forming a Quark-Nova.
}

   \authorrunning{Ouyed  et al. }            
   \titlerunning{SN 2009ip and SN 2010mc as dual-shock  Quark-Novae}  

   \maketitle


\section{Introduction}

SN 2009ip was first discovered as a candidate supernova (SN) by \citet{maza_2009}.  It was later shown consistent with Luminous blue variable (LBV) type behaviour \citep{miller_2009, li_2009, berger_2009} and dubbed an ``SN imposter" as over the next three years SN 2009ip went through a series of explosions resulting in a re-brightening by as much as 3 magnitudes in the R-band \citep{smith_2011, pastorello_2013}.  In early August 2012, the light curve (LC) of SN 2009ip increased to M$_{R}\sim$-15, brighter than any other outburst, and subsequently decayed over the next 40 days.  On September 23 2012, SN 2009ip re-brightened a final time, peaking at M$_{R}\sim$-18 and followed a LC similar to type IIn supernovae \citep{smith_2013a}.  The August event (2012a) followed by the September event (2012b) of SN 2009ip  was the clearest evidence of a double-hump in the LC of a SN so far reported.

SN 2010mc was discovered by \citet{ofek_2013} and also exhibited a pre-cursor outburst (2010a) $\sim$40 days before the main type IIn event (2010b).  \citet{smith_2013b} was the first to comment on the remarkable similarity between the SN 2009ip and SN 2010mc events, both in terms of LC and spectral evolution.  However, as far as we know, no pre-SN outbursts were observed in the years prior to the SN 2010mc event as was the case in SN 2009ip.  Owing to their uncanny similarity it is natural to conclude that SN 2009ip and SN 2010mc undergo similar processes at the end of their lives.

A debate on what to make of the SN 2009ip events in 2012 (and by association SN 2010mc) is currently under-way in the literature.  Several theories on the nature of the double-hump in the LC have emerged over the past year either claiming that the 2012 event was a true core-collapse SN, or simply more intense outbursts like those seen in the three years prior.  Several groups (e.g \citet{mauerhan_2013, smith_2013a, prieto_2013}) advocate that the first event (2012a) was a core-collapse SN while the second (2012b) was the interaction of the SN ejecta with the CSM.  \citet{fraser_2013} and \citet{pastorello_2013} argue in favour of the pulsational pair instability (PPI) mechanism in which the double-hump is explained through colliding shells of ejecta caused by two separate PPI explosions. \citet{margutti_2013} also support a two explosion scenario, concluding that the 2012b event is caused by the shock of the second explosion interacting with the material ejected from the first.  However, they do not come to a consensus on the nature of the two explosions.  Finally, the binary merger hypothesis was put forth by \citet{soker_2013} and \citet{kashi_2013} in which the multiple outbursts of SN 2009ip are caused by interaction of the binary system at periastron.  The final 2012b event in their model was the merger of the two stars in what they dub a ``mergerburst".

In this paper we present an alternative explanation for the double-hump in the LC of SN 2009ip  and SN 2010mc; the dual-shock Quark Nova (dsQN).  This two-explosion scenario was first put forth to explain the LC of SN2006gy \citep{leahy_2008, ouyed_2012} and later successfully applied to other super-luminous SN (SLSN) such as SN2005ap, SN2006tf, SN2007bi, SN2008es, SN2008fz, PF09cnd, PTF10cwr, SN2010gx \citep{kostka_2012, ouyed_2009} and SN2006oz \citep{ouyed_2013}; the latter showing evidence of a double-hump.  The double-hump is a key feature of the dsQN model and it was predicted in 2009 that future observations of SNe would show this in their LC \citep{ouyed_2009b}.  The paper is organized as follows: in \S 2 we give an overview of the dsQN model, in \S 3 we show the results of applying the dsQN model to SN 2009ip  and SN 2010mc.  We end with a discussion and conclusion in \S 4.

\section{Quark Nova Model}

The Quark Nova (QN) is the explosive event resulting from the transition of a neutron star to a quark star\footnote{The recent observations of a $\sim 2M_{\odot}$  NS \citep{demorest09}  does not rule out  the existence of quark stars.
Heavy quark stars may exist, so long as the strong coupling corrections are taken into account \citep{alford07}.
Furthermore, neutron stars and quark stars can co-exist since   not all neutron stars will be converted by the  capture of cosmic-ray strangelets
(\cite{bauswein09}).  Combining their simulations of strange star binary mergers with recent estimates of stellar binary populations, \cite{bauswein09} conclude that an unambiguous detection of an ordinary neutron star would not rule out the strange matter hypothesis.} \citep{ouyed_2002}; see \cite{ouyed_2013} for a recent review.    It was suggested  that this conversion, combined with the ensuing core collapse of the neutron star would result in an explosion causing on average $M_{\rm QN} \sim 10^{-3}$ M$_\odot$ of neutron-rich material to be ejected with $>$ $10^{52}$ erg of kinetic energy \citep{keranen_2005, ouyed_2009, niebergal_2010}.

In core-collapse SNe, neutrinos  carry away  99\% of the starÕs binding energy and drive the explosion. In QNe, neutrinos emitted from the quark core have  diffusion timescales  exceeding $\sim$ 10 ms (e.g. \cite{keranen_2005}) and cannot escape before the entire star converts to strange quark matter. 
For a QN, the agent of explosion is therefore photons, since the temperature of the quark core is large enough at the time of formation (much above the quark plasma frequency) to sustain large photon emissivities (\cite{vogt04}).   The mean free path is small enough to thermalize these photons inside the quark core, and  in the hadronic envelope so that energy deposition by photons is very efficient. For temperatures of  $\sim$ 1-10's MeV, the photon flux is a few  orders of magnitude higher than the neutrino flux from hot quark matter. The energy deposition in the NS outer layers (including the crust) is therefore much more efficient for photons than neutrinos and allows for ejecta with kinetic energy easily exceeding $10^{52}$ erg.

The fact that a few percent of the gravitational and conversion (from neutrons to quarks) energy is released as photons is unique to the QN \citep{ouyed05}. Even a few percent of the photon energy, when deposited in the thin crust of the neutron star, will impart a large momentum to it, leading to strong and ultra-relativistic mass ejection \cite{ouyed_2009}. The fate of this relativistic ejecta (with an average Lorentz factor $\Gamma_{\rm QN}\sim 10$) leads to a variety of observables including gamma-ray bursts (GRBs; e.g. \citep{ouyed_2011b}.), soft-gamma repeaters (SGRs; \cite{ouyed_2007a}), anomalous x-ray pulsars (AXPs; \cite{ouyed_2007b}), SNIa imposters \citep{ouyed_staff_2013}, r-process elements \citep{jaikumar_2007}, SLSNe and double-humped SNe (e.g. \citep{ouyed_2009b}).

\subsection{Dual-Shock Quark Nova Model}

The dsQN happens when the QN occurs days to weeks after the initial SN, allowing the QN ejecta to catch up to and collide with the SN remnant.  Shock reheating occurs at a large radius (because of the time delay between QN and SN) so that standard adiabatic losses inherent to SN ejecta are far smaller. Effectively, the SN provides the material at large radius and the QN re-energizes it, causing a re-brightening of the SN.  For small time delays ($\sim$days) the radius of the SN ejecta is relatively small resulting in a modest re-brightening when the QN ejecta catches up.  In this case the re-brightening may occur during the rise of the initial SN and be hidden from direct observation, although unique spallation products may be identified in the spectrum \citep{ouyed_2011b}. If the time-delay is $\sim$a few weeks, the radius and density of the SN ejecta will be optimal for an extreme re-brightening as observed in SLSNe \citep{leahy_2008, ouyed_2012, kostka_2012, ouyed_2013}.  If a QN goes off in isolation (i.e. time-delays $> \sim$ a few months), the SN ejecta will be too large and diffuse to experience any re-brightening by the QN ejecta.

It is clear that if the timing is right, and the re-brightening is not buried in the SN LC, a double-hump in the LC should be observed.  The first hump corresponds to the SN whereas the second is the re-brightening of the SN ejecta when it is hit by the QN ejecta (see Figs. 2 and 3 in \citet{ouyed_2009b}).  For time-delays $\sim$ a month, the second hump is expected to be similar in brightness to the first (i.e. not super-luminous).  The peak of the second hump occurs when the shock breaks out of the SN ejecta at $t_{\rm sbo} = t_{\rm delay} + t_{\rm prop}$ where $t_{\rm delay}$ is the time-delay between the SN and QN explosions and $t_{\rm prop}$ is the time for the QN ejecta to catch up to and for the resulting QN shock to propagate through the SN ejecta;  the relativistic QN ejecta catches up to the SN ejecta on very short time scale, $\sim (v_{\rm SN}/c) t_{\rm delay}$ where $c$ is the speed of light.

\section{Results}

We fit the LCs of SN 2009ip  and SN 2010mc using a three component model: the SN, dsQN and wind. The SN and QN models are those used in \citet{leahy_2008}.  The key parameters in the SN model are the radius of the progenitor star ($R_{\star} = 30 M_\odot$), the mass of the SN ejecta ($M_{\rm SN}$), the energy of the SN ($E_{\rm SN}$) and the velocity of the SN ejecta ($v_{\rm SN}$).  The additional parameters of the dsQN model include the time-delay between SN and QN ($t_{\rm delay}$), the velocity of the QN shock  ($v_{\rm QN, shock}$) and the energy of the QN explosion ($E_{\rm QN} = 10^{52}$erg).

The LCs show clear evidence of emission beyond the two humps which is explained in our model as the collision of the combined SN/QN ejecta with the surrounding CSM. To this effect,  we use the analytical bolometric light curve model of \citet{moriya_2013} since these models are shown to agree well with numerical light curves. These models assume a constant CSM (i.e. wind) velocity $v_{\rm w}$ and a CSM density profile $\rho_{\rm CSM} = D r^{-s}$ where $D$ is a constant. The corresponding mass-loss rate is  $\dot{M}_{\rm w} = \rho_{\rm  CSM} v_{\rm w} 4\pi r^2 = D v_{\rm w} 4\pi r^{2-s}$ with the $s=2$ case  corresponding to the steady mass-loss scenario  where $D= \dot{M}_{\rm w}/4\pi v_{\rm w}$. The combined SN/QN ejecta is defined by its kinetic energy $E_{\rm SNQN}$ and its mass $M_{\rm SNQN}=M_{\rm SN}$ ($M_{\rm QN} << M_{\rm SN}$). Its has a double power-law profile for the density of homologously expanding ejecta ($\rho_{\rm ej}\propto r^{-n}$ outside and $\rho_{\rm ej}\propto r^{-\delta}$ inside). Another parameter in these models is the conversion efficiency from kinetic energy to radiation, $\epsilon$. We assume an inner radius for the CSM which is equivalent to a time delay for the CSM interaction after SN. We call this time delay $t_{\rm CSM}$. For each fit, we use $n=7$, $\delta = 0$ and $\epsilon = 0.1$. 
 We refer the interested reader to \S~2.2 in \citet{moriya_2013} for the equations and parameters we used in our modelling of the CSM interaction.

 We fit the three component model (SN, dsQN and wind) to the observations
by computing models for a variety of parameters until we found parameters
that gave a combined lightcurve which agreed well with the observations.
The time required to generate a model does not allow for an efficient high
precision parameter search, and so the parameters represent an good
manually-obtained fit rather than ``best fit" from minimizing $\chi^2$.

\subsection{SN 2010mc}

The LC fit of SN 2010mc using the three component model described above is shown in Figure \ref{fig:fig_1} along-side the observations from Ofek et al. (2013).  The SN and dsQN models fit the first two humps well and the CSM model fits the late-time LC from $\sim$ day 130 onwards.  To produce the fit, we required  $M_{\rm SN} = 20 M_{\odot}$, $v_{\rm SN} = 3000$ km~s$^{-1}$, $t_{\rm delay} = 33$ days,  and $v_{\rm QN, shock} = 8000$ km~s$^{-1}$.  We also required a wind velocity of $v_{\rm w} = 1000$ km~s$^{-1}$ leading to a modest mass loss rate for the wind of $\dot{M}_{\rm w} \sim 5 \times 10^{-7}$ M$_\odot$ yr$^{-1}$.

\subsection{SN 2009ip }

\begin{figure*}[t!]
\resizebox{\hsize}{!}{\includegraphics{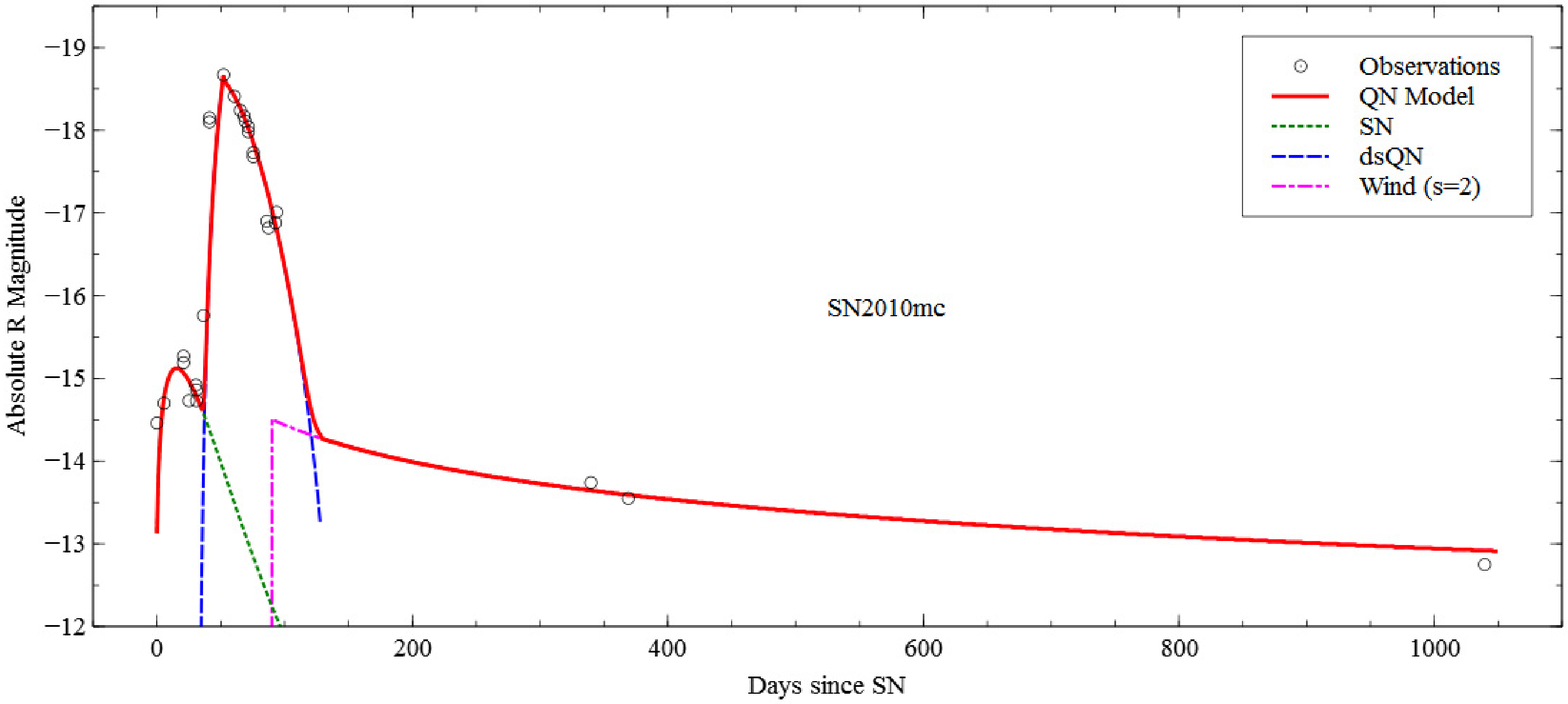}}
\caption{dsQN model fit (red solid line) to the light curve of SN 2010mc.  The observations (black open circles) are from \citet{ofek_2013}.  The green dotted line represents the SN light curve, the blue dashed line the interaction between the QN and SN ejecta and the magenta dashed-dot line represents the interaction between SN/QN ejecta and the CSM wind.}
\label{fig:fig_1}
%
\resizebox{\hsize}{!}{\includegraphics{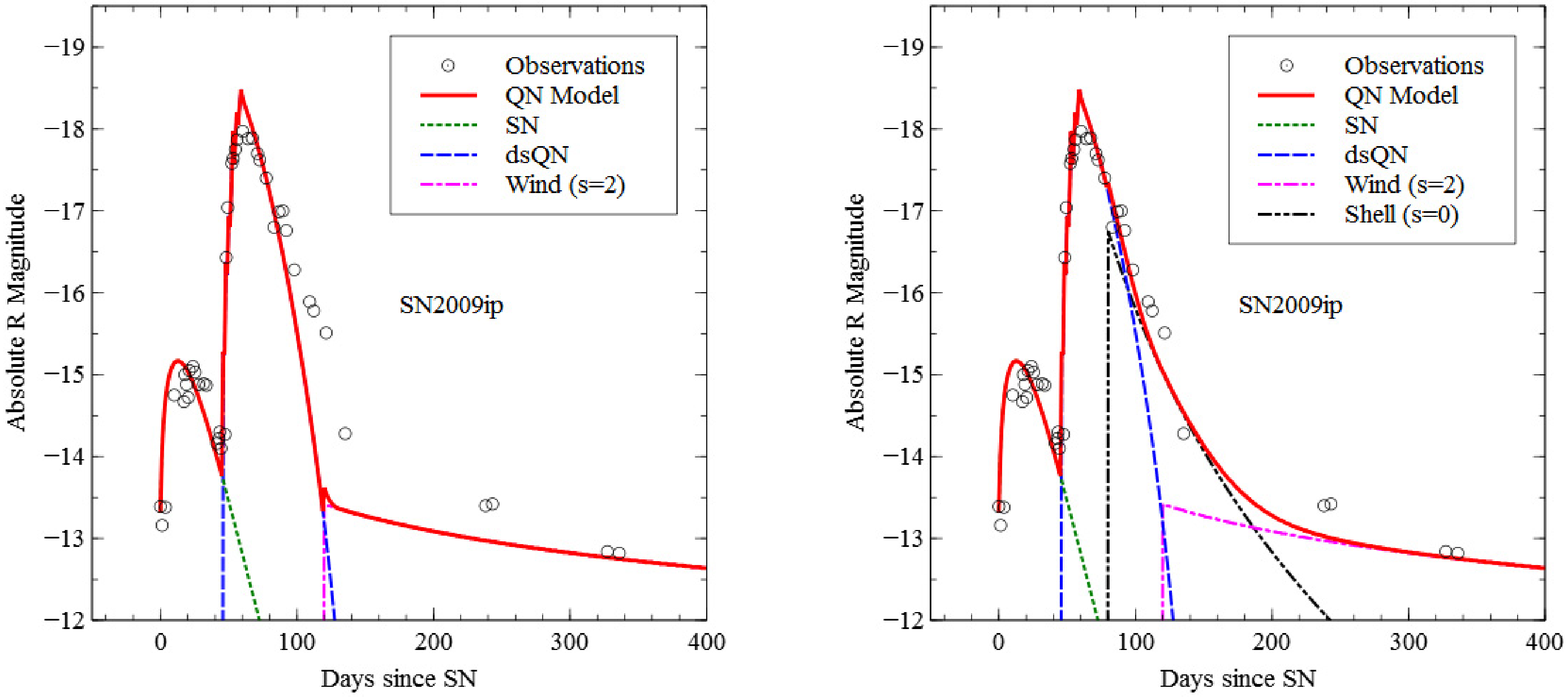}}
\caption{dsQN model fit (red solid line) to the light curve of SN 2009ip .  The observations (black open circles) are from \citet{smith_2013a}.  The green dotted line represents the SN light curve, the blue dashed line the interaction between the QN and SN ejecta and the magenta dashed-dot line represents the interaction between SN/QN ejecta and the CSM wind.  The left panel is a 3-component fit to the data whereas the right panel shows a fit using an additional CSM shell component (black dot-dashed line).}
\label{fig:fig_2}
\end{figure*}

We fit the LC of SN 2009ip  using the three-component model in the left panel of Figure \ref{fig:fig_2}.  The observations plotted along-side our fit are from \citet{smith_2013a}. The parameters used in this fit are $M_{\rm SN} = 25 M_{\odot}$, $v_{\rm SN} = 3600$ km~s$^{-1}$,  $t_{\rm delay} = 40$ days, $v_{\rm QN, shock} = 11000$ km~s$^{-1}$ and $v_{\rm w} = 2000$ km~s$^{-1}$ leading to $\dot{M}_{\rm w} \sim 2 \times 10^{-7}$ M$_\odot$ yr$^{-1}$.  The first and second hump are fit well by the SN and dsQN model, as in the case of SN 2010mc, but the CSM wind model fails to account for the late-time LC until $\sim$ day 300.  This implies the presence of a fourth component, active between $\sim$day 80-300.  SN 2009ip  went through several violent explosions in the three years prior to the 2012a event resulting in (presumably) dense shells of CSM.  Our fourth component is therefore the interaction of the SN/QN ejecta with one of these dense shells.  We model the shell using the same prescription above, only with $s=0$ instead of $s=2$.  The right panel of Figure \ref{fig:fig_2} shows the result of adding this fourth component to the model which significantly improves the model fit.  The parameters for the SN 2009ip  and SN 2010mc fits are summarized in Tables \ref{table:double-hump} and \ref{table:late-time}.

\section{Discussion and Conclusion}

In our model, the initial hump observed in the LC of SN 2009ip  and SN 2010mc is caused by the SN explosion.  The second hump is the re-brightening caused when the QN ejecta catches up to and collides with the SN ejecta and the LC tail is a consequence of the SN/QN material interacting with the CSM wind.

Using the dsQN model, we have been able to account for the double-hump in the LC of both SN 2009ip  and SN 2010mc.  The two humps are fit with very similar parameters, owing to their remarkably similar LCs.  The late time LC of SN 2009ip  is not fit well with a single CSM wind model and requires the interaction with a CSM shell of constant density.  SN 2009ip  went through several LVB-like outbursts in the years prior to the 2012 event so the existence of these shells are expected.  For a shell velocity of $\sim600$ km s$^{-1}$ \citep{smith_2013a}, $v_{\rm SN}=3600$ km s$^{-1}$ and a $t_{\rm CSM}=80$ days (both fit parameters from this work) we can infer that the shell was ejected $\sim400$ days prior to the SN explosion.  This places the mass ejection event $\sim~$July 2011, right in the middle of the 2011 outburst \citep{pastorello_2013}.  Given the multiple outbursts prior to 2012, it is likely that many such shells exists \citep{margutti_2013}.  We choose to only include one interaction in this work to illustrate the idea, although adding more would certainly improve the fit to the LC (and increase the number of free parameters).  We should note that we did not need a shell component in our model of SN 2010mc, suggesting that it did not suffer violent mass ejections prior to the 2010 event on the same level as SN 2009ip .  This seems to be supported by the fact that no pre-burst activity was detected for SN 2010mc (although absence of proof is not proof of absence).  We therefore suggest that the progenitor of SN 2010mc was not necessarily an LBV star (as is the general consensus for SN 2009ip ), but could have been an O-star.  This is consistent with the QN scenario, in that the type of progenitor is not a concern, as long as it explodes with a low enough mass to leave behind a massive neutron star (i.e. a QN progenitor) rather than a black hole.

\citet{margutti_2013} argues that a double explosion is needed to explain the observations of SN 2009ip , while \citet{smith_2013a} provides strong evidence that the 2012a event was a SN.  Most authors agree that the tail of the LC is supported by interaction between ejecta.  \cite{martin_2013}  find significant fluctuations during the decline of  SN 2009ip lightcurve
past the main peak. They argue against the CSM interaction hypothesis and suggest that
these fluctuations may be caused by the progenitor of SN 2009ip which survived its last outburst.
In our model, where the CSM interaction comes after the double-hump, these bumps are indicative of the SN/QN material interacting with the different shells ejected during the pre-2012 eruptions. It is natural, in this context, that the fluctuations are consistent with those detected before 2012a since they are essentially an ``explosion" echo of the previous bursts. These bumps should be observed in the lightcurve of all dsQNe where the progenitor was an LBV star with relatively recent eruptions.

Our dsQN model provides a natural unifying framework which compliments each of these points.  Further, the similarity between the LCs of SN 2009ip  and SN 2010mc is simply a consequence of their similar time-delays.  No fine tuning of parameters is needed to explain their coincidence.
The key ingredient is a progenitor in the right mass range to produce a massive enough NS but not a black hole (i.e. a progenitor leading to a SN ejecta with $M_{\rm SN}\sim 20M_{\odot}$ according to our findings).  

In summary, we have learned that regardless of the nature of the star (O-star, LBV-star etc.), if the mass of the ejected envelope in the SN is $\sim 20M_{\odot}$, the core seems to meet the conditions to produce a NS massive enough to experience a QN explosion within weeks.  These different stars would lead to a variety of SNe (the first hump) and late time emission.  The brightness and shape of the second (i.e. QN) hump depends mainly on the time delay between the SN and QN (Leahy\&Ouyed 2008; Ouyed et al. 2009). 

With so many possible theories being presented in the literature, how can we prove/disprove the dsQN hypothesis?  The QN has several unique signatures that are not expected in other models: 

\begin{itemize}

\item Aligned rotator: The conversion of neutron star to quark star aligns the magnetic field with the axis of rotation
\citep{ouyed_2004,ouyed_2006}.  A radio pulsar  is therefore not expected at the location of SN 2009ip and SN 2010mc. 
  However, if beaming is favorable, a parent radio pulsar  should be detectable  in
the period between the SN event and the QN event, particularly once the SN hump is on the decline.

\item AXPs/SGRs:  If some of the QN ejecta fell back and remained in orbit around the quark star, we would expect to see AXP/SGR behaviour around SN 2009ip  and SN 2010mc \citep{ouyed_2007a,ouyed_2007b}.

\item  r-Process elements: The QN provides the ideal site for the creation of  r-process elements with atomic weight $A>130$ \citep{jaikumar_2007}.  We should therefore expect to see evidence of these heavy elements in the late-time spectra of SN 2009ip  and SN 2010mc.

\item Neutron decay:  Since the QN is an explosion of a neutron star, a large fraction of the ejecta is composed of free neutrons.  Free neutrons decay in $\sim$900 seconds (longer if they are relativistic) with unique electromagnetic signatures (see e.g. \citet{severijns_2006, nico_2006}).  We therefore expect to see a release of energy soon ($\sim$hours) after the QN explosion.  The exact nature of this signature is still unknown, but may occur at energies $>$ 15~keV (assuming it is not absorbed by overlying SN material).

\item Gravitational waves:  In the QN scenario, there are two violent explosions (the SN and QN) that will give distinct gravitational wave signatures \citep{staff_2012}.  Future gravitational wave observations of a SN exhibiting a double-hump LC could shed light on the explosion mechanism.

\item    The first hump (the SN) should show signatures of typical 
SN r-process elements (e.g. \cite{takahashi04}) while the second hump should
include much heavier r-processed elements  (with $A>130$) deposited by the QN ejecta (Jaikumar et al. 2007). 
The QN ejecta is of the order of $10^{-3}M_{\odot}$  which should yield heavy elements in amounts exceeding the 
$10^{-6}M_{\odot}$ values   processed in a typical  SN.

\end{itemize}

\begin{acknowledgements}   

This work is funded by the Natural Sciences and Engineering Research Council of Canada. N.K. would like to 
acknowledge support from the Killam Trusts.

\end{acknowledgements}



\begin{table*}[t!]
\caption{Parameters used for fitting double-hump light curve of SN 2009ip  and SN 2010mc}
\begin{center}
\begin{tabular}{|c|c|c|c|c|c|}
\hline
& \multicolumn{3}{c|}{SN parameters} & \multicolumn{2}{c|}{QN parameters}\\
\hline
Source & $M_{\rm SN}$($M_{\odot}$) & $E_{\rm SN}$ ($10^{50}$ ergs)  &  $v_{\rm SN}$ (km s$^{-1}$) & $t_{\rm delay}$ (days)   &  $v_{\rm QN, shock}$  (km s$^{-1}$)\\
\hline

SN 2009ip  & 20 & 2.0 & 3600 & 40 & 11000\\
\hline
SN 2010mc & 25 & 4.5 & 3000 & 33 & 8000\\
\hline

\end{tabular}
\end{center}
\label{table:double-hump}
%
\caption{Parameters used for fitting late-time light curve of SN 2009ip  and SN 2010mc}
\begin{center}
\begin{tabular}{|c|c|c|c|c|c|c|c|}
\hline
Source  &   D  & $s$ &  $t_{\rm CSM}$   (days)  & Possible Progenitor \\ \hline 
SN 2009ip (wind) & $5.0 \times 10^{9}$ g cm$^{-1}$ &  2 & 120 & \multirow{2}{*}{LBV star} \\
\cline{1-4} 
SN 2009ip (shell) & $1.0 \times 10^{-9.3}$ g cm$^{-3}$ & 0 & 80 &\\
\hline \hline
SN 2010mc (wind) & $2.5 \times 10^{10}$ g cm$^{-1}$ & 2 & 90 & O-star\\
\hline 
\end{tabular}\\
~\\~\\~\\
\end{center}
\label{table:late-time}
\end{table*}


\end{document}